\documentclass[twocolumn,showpacs,prl,amsmath,amssymb,floatfix,superscriptaddress]{revtex4}
\usepackage{dcolumn}
\usepackage[pdftex]{graphicx}
\usepackage{color}
\begin{document}
\newcommand{\blue}{\textcolor{blue}}

\title{Dominant parameters for the critical\\tunneling current in bilayer exciton condensates}

\author{L. Tiemann\footnote{Correspondence should be send to: lars@will.brl.ntt.co.jp}}
\author{Y.Yoon}
\author{W.Dietsche}
\author{K. von Klitzing}
\affiliation{Max-Planck-Institut f\"{u}r Festk\"{o}rperforschung, Heisenbergstra{\ss}e 1, D-70569 Stuttgart, Germany}
\author{W. Wegscheider}
\affiliation{Institut f\"{u}r Experimentelle und Angewandte Physik, Universit\"{a}t Regensburg, D-93040 Regensburg, Germany}%

\begin{abstract}
We will discuss the relevant conditions to observe a critical tunneling current [New J. Phys. 10, 045018 (2008)] in electron-double layer systems at a total filling factor of one and find they are related to the effective layer separation and the temperature. Our studies suggest that the intensity of the critical tunneling behavior is also directly linked to the area of the sample.
\end{abstract}

\maketitle 

\section{I. Introduction}

Under large perpendicular magnetic fields $B$ the motion of charged carriers in two-dimensional electron systems (2DES) is confined to small cyclotron orbitals. This confinement suppresses the kinetic energy of the electrons but on the other hand amplifies their Coulomb interactions. In single layers these Coulomb correlations can lead to the emergence of fractional quantum Hall states. Two individual but closely-spaced 2DES may also exhibit a correlated state, however, the underlying physics is now also influenced
by the Coulomb interactions between the two systems. When the electron densities $n$ in both layers are identical and the individual filling factors $\nu=nh/eB$ are close to 1/2 (i.e., $\nu_{tot}=1$), the system may spontaneously develop interlayer phase coherence, provided the distance $d$ between the layers is sufficiently small. This led to the prediction of Josephson-type phenomena in bilayer systems \cite{Fertig89, Ezawa92, Wen92} nearly 20 years ago \footnote{The underlying notion of electron-hole pairing in semiconductors and its condensation relies on the pioneering works of M. Blatt, K. W. B\"{o}er and W. Brandt [Phys. Rev. 126, 1691 (1962)], L. V. Keldysh and A. N. Kozlov [Zh. Eksp. Teor. Fiz. 54, 978 (1968)], Yu. E Lozovik and V.I. Yudson [JETF Lett. 22, 11 (1975)] and S. I. Shevchenko [Fiz. Nizk. Temp. 2, 505 (1976); Sov. J. Low Temp. Phys. 2, 251 (1977)].}. The ratio of the center-to-center layer separation $d$ and the magnetic length $l_B=\sqrt{\hbar /eB}$ is commonly used to parameterize the strength of this emerging state at the bilayer's total filling factor of one. Tunneling spectroscopy experiments \cite{Spielman00, Spielman01} demonstrated the phase coherence between the two layers by showing a dramatic enhancement of the tunneling conductance at $\nu_{tot}=1$. More recently, a critical behavior was observed in dc tunneling experiments \cite{Tiemann08b}, which had also been predicted \cite{Ezawa94} but failed to appear in all prior experiments. The purpose of this paper is to elucidate on the requirements to observe such a critical behavior. Our studies demonstrate that the coherent tunneling not only intensifies with the size of the sample but more importantly that the critical current grows linearly with the $\nu_{tot}=1$ area. In order to deliver a thorough picture of the critical tunneling behavior we first briefly address those experimental conditions which can easily be manipulated, such as effective layer separation $d/l_B$, temperature and filling factor, before discussing the more relevant size dependence.


\section{II. Experimental Details}
\label{secexpdetails}

\begin{figure*}[!htp]
\centering
 \includegraphics[width=0.99\textwidth]{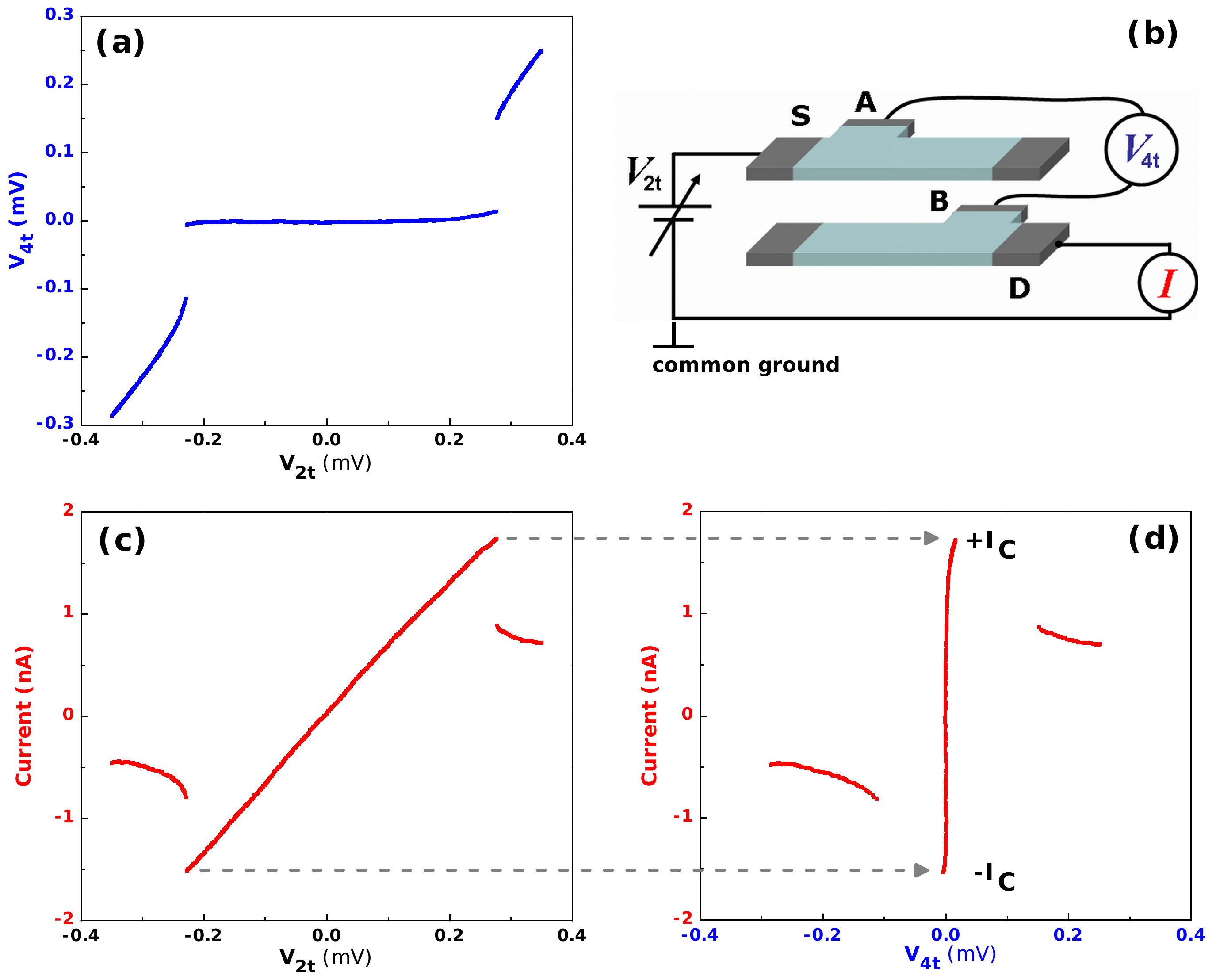}
 \caption{\textbf{(a) Measured 4-terminal voltage $V_{4t}$ (c) and the measured total current $I$ as a function of the applied 2-terminal voltage $V_{2t}$ at $\nu_{tot}=1$ ($d/l_B$=1.42). When the current is plotted over $V_{4t}$, the curve (c) collapses onto subfigure (d). The absolute values of the negative and positive critical currents $\pm I_C$, however, remain unchanged and translate now into the maximal currents around $V_{4t}=0$. Subfigure (b) is a cartoon of the measurement setup with S and D being the source and drain contacts and A and B the voltage probes. Hall bar sample of 0.88~$\times$~0.08~mm$^2$ (wafer $\alpha$) at T$<$20~mK.}} \label{fig1}
\end{figure*} 

Our data are obtained from samples of two different wafers grown in two different molecular beam epitaxy (MBE) machines (hereby referred to as wafer $\alpha$ and $\beta$), both with a net barrier thickness of 9.6~nm, consisting of alternating AlAs (1.70~nm) and GaAs (0.28~nm) layers. Wafer $\alpha$ is the same as used in \cite{Tiemann08b}. The intrinsic densities of approximately 4.0~$\times~10^{10}$~cm$^{-2}$ (wafer $\alpha$ and $\beta$) of the two quantum wells originate from standard modulation doping. The low-temperature mobilities exceed 450000~cm$^{2}$V$^{-1}$s$^{-1}$ (wafer $\alpha$) and 500000~cm$^{2}$V$^{-1}$s$^{-1}$ (wafer $\beta$). The single-particle tunnel splitting $\Delta_{S,AS}$ in our double quantum wells was estimated to be approximately 150~$\mu$K. Independent electrical contact to the two layers is achieved by growing the double quantum wells onto prestructured back gates \cite{Rubel97} and by using additional top gates in order to exploit a selective depletion technique \cite{Eisenstein90}. As the distance between back gate and 2DES is only around 1~$\mu$m, this overgrown back gate technique requires voltages of less than 1~V to locally deplete the contact arms and obtain independent electrical contacts to the two layers. The samples were either patterned into Hall bars of different sizes or a quasi-Corbino ring \cite{Tiemann08a}. The specific sample dimensions will be given later in the text.


\begin{figure}[!htp]
\centering
 \includegraphics[width=0.49\textwidth]{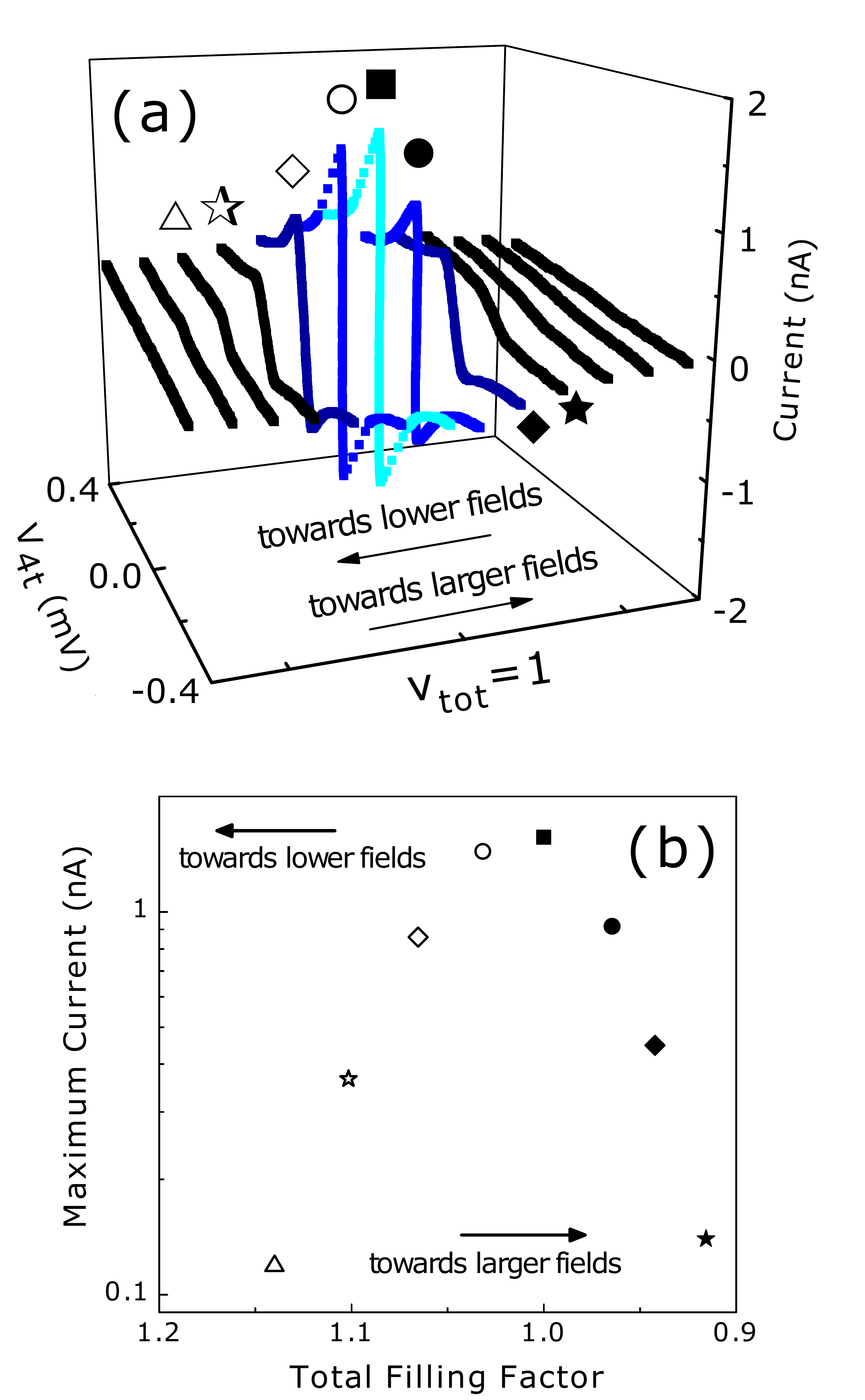}
 \caption{\textbf{(a) Tunnel characteristics for several total filling factors, i.e., the electron density remains constant while $I/V$ characteristics are measured at several magnetic fields around $\nu_{tot}=1$. The eight most inner curves are marked with symbols. (b) Maximal (critical) currents as a function of the total filling factor (data from labeled curves in subfigure (a) are used). At exactly $\nu_{tot}=1$ the effective layer separation is 1.42. Hall bar sample of 0.88~$\times$~0.08~mm$^2$ (wafer $\alpha$) and $T_{bath}<20~$mK.}} \label{fig4}
\end{figure}

The dc $I/V$ tunneling measurements presented in this paper are performed as sketched in Figure \ref{fig1} (b), by applying a tuneable dc bias voltage (herein after referred to as 2-terminal voltage V$_{2t}$) between the two layers and detecting the current flow $I$ toward ground as a voltage drop over a known resistance \cite{Tiemann08b}. As the interlayer phase coherence at $\nu_{tot}=1$ allows to easily transfer charges between the layers, the tunnel conductance becomes enormously enhanced which is tantamount to a very small tunnel resistance. Consequently, even if there is a finite bias $V_{2t}$ applied, the interlayer voltage probes A and B (located close to source and drain) may read a value for $V_{4t}$ of close to zero (Figure \ref{fig1} (a)). However, this coherent tunneling can be destroyed if the current (Figure \ref{fig1} (c)) exceeds a critical value $I_C$. For the representative measurement presented in Figure \ref{fig1} the critical value is roughly $\pm$1.5~nA. Figure \ref{fig1} (d) is a depiction of the measured current $I$ over the measured voltage $V_{4t}$. This representation will be used throughout this paper, and the critical currents $\pm I_C$ now translate into the maximal positive and negative currents at V$_{4t}$=0 as indicated by the dashed lines. The data here are presented in a scatter plot which show a discontinuity in the measured current and voltage characteristics when the system moves from the coherent strong tunneling regime into the weak tunneling regime. The origin of this negative differential conductance is the sudden change in the total impedance $R_{tot}=R_T+R_S$, when a tunneling resistance $R_T$ of almost zero is replaced by a resistance much larger than the series resistance $R_S$ of the (uncorrelated) quantum Hall systems, i.e., $R_T\approx 0~\Omega \rightarrow R_T \gg R_S$.

Historically, the study of coherent tunneling at $\nu_{tot}=1$ exploits the tunneling-spectroscopy technique where the differential tunneling conductance $dI/dV$ is obtained in an ac measurement. These tunneling-spectroscopy measurements (TSM) reveal a resonantly enhanced zero bias tunneling peak at a total filling factor of one \cite{Spielman00, Spielman01, Wiersma06, Finck08, Champagne08}. A critical behavior as discussed in this paper, however, is hidden in the dc part of the TSM which is usually not shown. If the critical current is intrinsically small, it may also be difficult to detect or conceal through the influence of the ac modulation. 

\begin{figure*}[!htp]
\centering
 \includegraphics[width=0.99\textwidth]{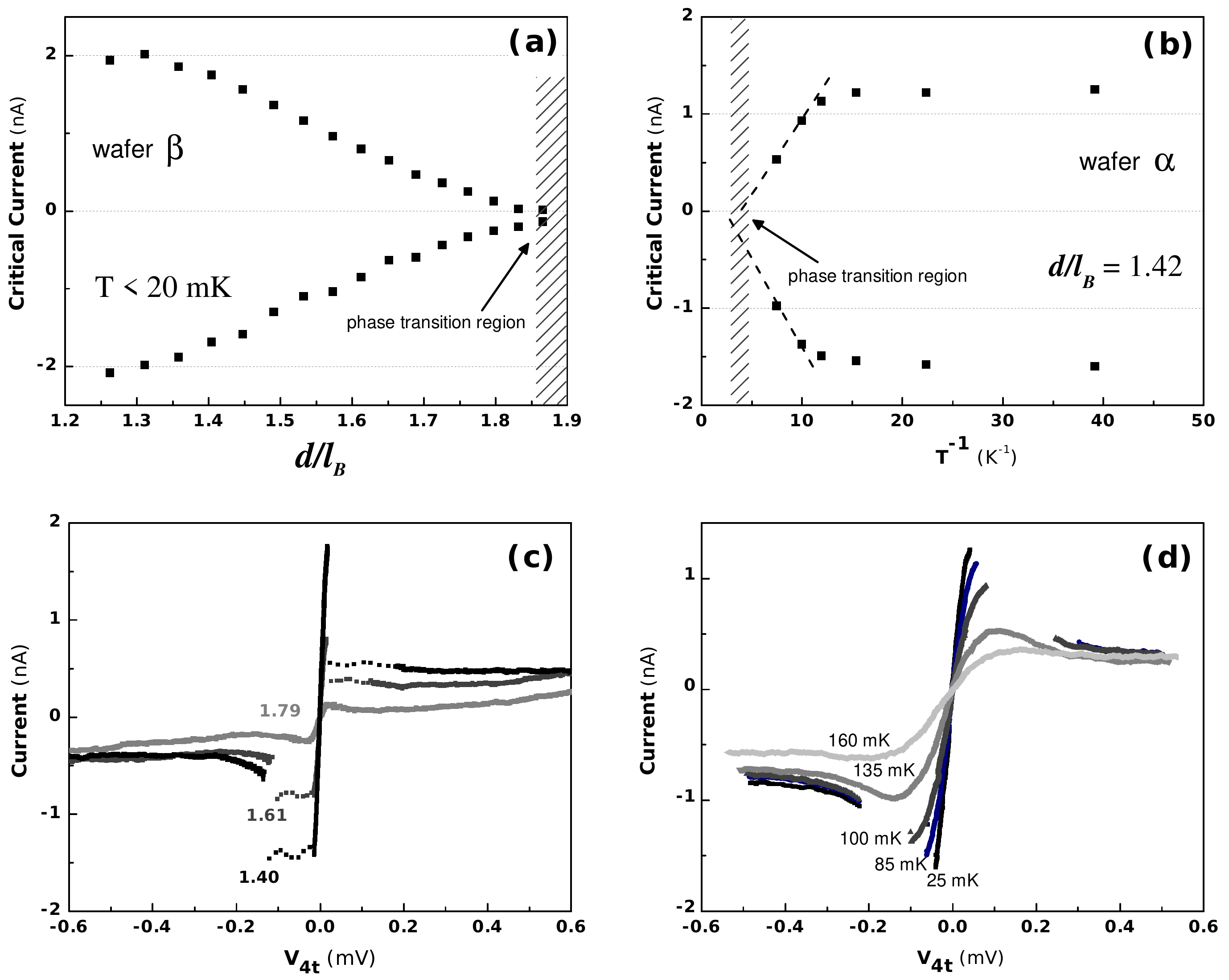}
 \caption{\textbf{Critical current as a function of the effective layer separation $d/l_B$ at fixed temperature (a) and the inverse temperature at fixed effective layer separation (b). Figures (c) and (d) illustrate some corresponding 4-terminal $I/V$ curves to allow comparison.}} \label{fig3}
\end{figure*}


\section{III. Critical current vs. filling factor}

It does not come as a surprise that the critical tunneling behavior (or the signatures of the $\nu_{tot}=1$ state in general), is strongest when the total filling factor of the system is exactly 1, or when the individual filling factors are exactly 1/2, respectively \footnote {In this paper we remain in the matched density condition, i.e., $n_1$=$n_2$. However, imbalancing the electron densities can be used as a tool to induce and intensify the $\nu_{tot}=1$ state. See Spielman \textit{et al.}, Phys. Rev. B 70, 081303 (2004) and \cite{Wiersma04} for details.}. Deviating to larger fields (smaller total filling factors) or smaller fields (larger total filling factors) will strongly suppress the coherent tunneling. In the uncorrelated regimes, carriers are exchanged between two individual two-dimensional electron systems, a process which now requires a finite amount of energy to overcome the Coulomb repulsion between the electrons. Figure \ref{fig4} (a) provides a qualitative picture of this behavior. It shows a set of several 4-terminal $I/V$ curves measured on a Hall bar sample of 0.88~$\times$~0.08~mm$^2$ (wafer $\alpha$) at $\nu_{tot}=1$ ($d/l_B=1.42$) and for small offsets in steps of about $\Delta{\nu}=0.3$ thereof. While drifting to either side of $\nu_{tot}=1$, the critical tunneling behavior is getting progressively suppressed around zero bias. Our data suggest that moving toward higher fields suppresses the tunneling slightly more rapidly than in the opposite low field direction as the Coulomb exchange increases with $l_B^{-1}\propto B$. Figure \ref{fig4} (b) gives a quantitative analysis of the maximum (critical) current as a function of the total filling factor. Note that only the eight plots in the center of Figure \ref{fig4} (a) allowed the determination of a maximal current. 


\section{IV. Critical current vs. effective layer separation and temperature}

As the total filling factor one state resides within a large parameter space, the size dependence as we are about to discuss cannot be studied fully independent of other important parameters. The purpose of Sec. IV is thus to outline the effects of manipulating temperature and effective layer separation $d/l_B$ which largely influence the magnitude of the critical current as well. The latter is achieved experimentally by increasing the electron densities in both layers simultaneously and adjusting the magnetic field, i.e., $B_{v_{tot=1}}\propto n_{tot}$. Figure \ref{fig3} (a) shows the positive and negative critical currents as a function of $d/l_B$ for a Hall bar sample of 0.88~$\times$~0.08~mm$^2$ (wafer $\beta$) at below 20~mK. For $d/l_B \leq 1.3$ the current appears to saturate at 2~nA, even to decrease. This however is related to the effect of the sample's very low electron density, which begins to suppress the transport current altogether (the single layer density is roughly 1.65~$\times$~10$^{14}$~m$^{-2}$). For intermediate $d/l_B$ on the other hand the trend is clearly linear. For $d/l_B> 1.85$ the system undergoes a phase transition and the critical tunneling behavior (and the $\nu_{tot}=1$ QH state) disappears. This value is in very good agreement with those found in magneto-transport \cite{Murphy94} or tunneling-spectroscopy experiments \cite{Spielman00}. Generally, if the onset of the $\nu_{tot}=1$ state is observed at $d/l_B\approx 2$, its origin is a pure many-body effect, and in weakly tunneling samples (i.e., $\Delta_{S,AS}\approx 0$) the phase-coherence would develop spontaneously \cite{Murphy94}. The smooth phase transition we observed supports a ``puddle model'' as suggested by A. Stern and B. I. Halperin \cite{Stern02}
where the $\nu_{tot}=1$ phase breaks up into domains near the phase boundary. In this model, the type of phase transition would have to be of first order as two phases co-exist in the sample. Figure \ref{fig3} (c) shows three 4-terminal $I/V$ curves corresponding to a $d/l_B$ of 1.40, 1.61 and 1.79.

The expected phase transition for correlated bilayers occurring at finite temperatures is not a regular second order phase transition such as for normal superconductors at zero field or ferromagnets but a Kosterlitz-Thouless type of phase transition. However, standard transport experiments are not able to judge type and form of the occurring phase transition. Nevertheless, what we see experimentally in transport (i.e., tunneling) at finite temperatures is summarized in Figure \ref{fig3} (b). There we plot the critical current as a function of the inverse temperature for a fixed $d/l_B=1.42$ measured on a Hall bar sample of 0.88~$\times$~0.08~mm$^2$ (wafer $\alpha$). Figure \ref{fig3} (d) shows several corresponding 4-terminal $I/V$ curves. At a temperature exceeding approximately 80~mK, the critical current begins to decrease rapidly. Extrapolation of the data in this region indicates the suppression of the critical behavior for temperatures above 250~mK. The $\nu_{tot}=1$ QH state as observed in magneto-transport is very weak at 250~mK and disappears entirely at temperatures exceeding approximately 350~mK \cite{Tiemann08a}. The exact type of this phase transition is unknown to us. 
Spielman \cite{Spielman04} found similar overall trends on the temperature and $d/l_B$ in tunneling spectroscopy measurements using samples which display very small critical currents.


\section{V. Critical current vs. sample size}

The following study of the size dependence was motivated by magneto-transport and tunneling measurements performed on a Corbino ring \cite{Tiemann08a}. These experiments had shown a vanishing conductance across the annulus, suggesting that the bulk of the $\nu_{tot}=1$ phase is incompressible. As any interedge charge transfer is suppressed, tunneling would then (generally) only occur in the vicinity of the edges of the coherent $\nu_{tot}=1$ system and its magnitude would have to scale with the circumference of the sample. Tunneling spectroscopy measurements on Hall bar samples on the other hand indicated that the zero bias tunneling conductance may be related to the area of the sample instead \cite{Finck08}. To obtain a better understanding and trying to solve this contradiction we compared the critical currents, rather the properties of the TSM tunneling peak, in terms of the circumference and area. Comparing different samples of course may introduce a systematic error, yet we still found that the effect of different sample sizes had much more dramatic consequences as we will see next.

\begin{figure}[!htp]
\centering
 \includegraphics[width=0.49\textwidth]{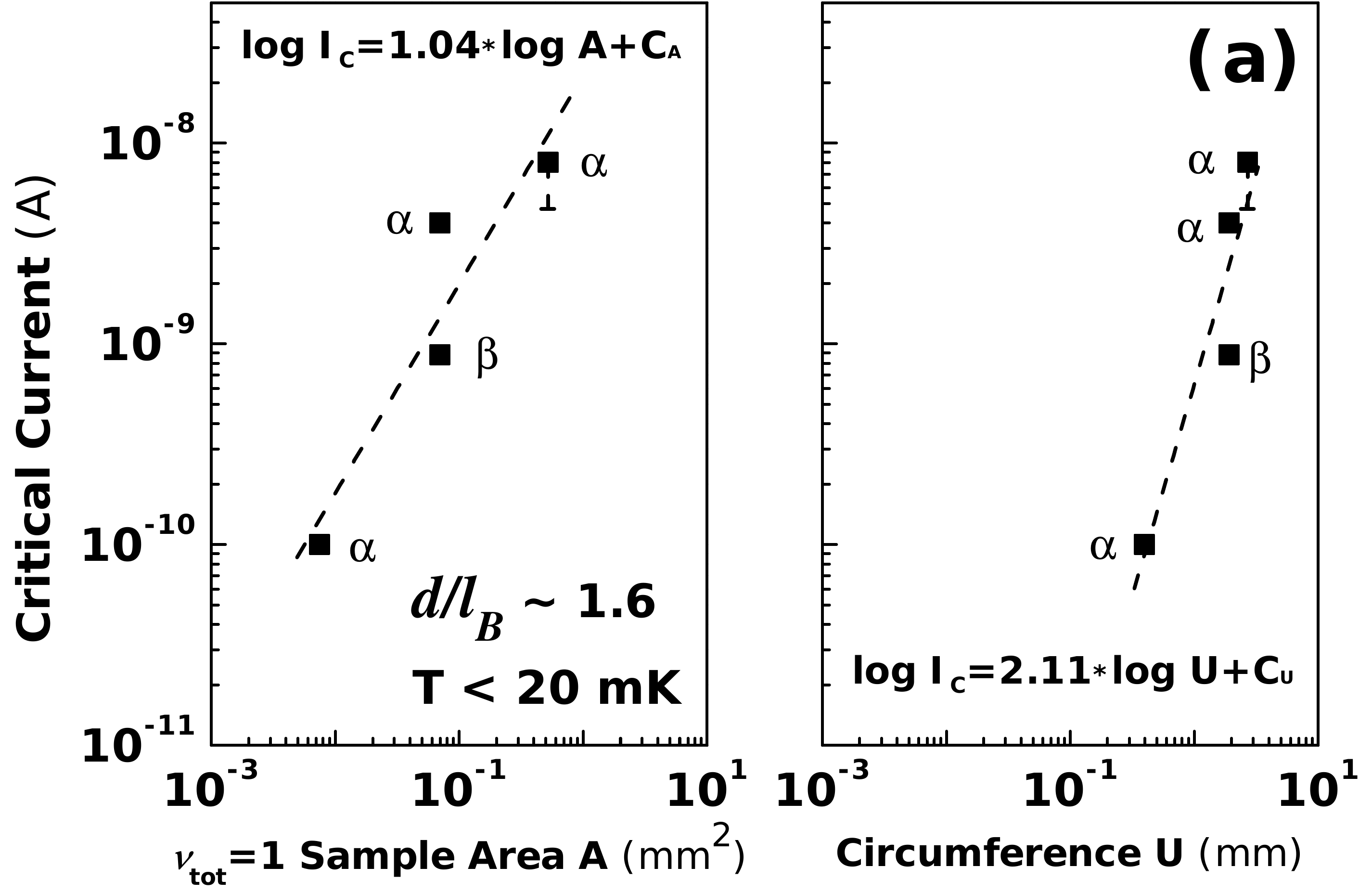}\\
 \includegraphics[width=0.47\textwidth]{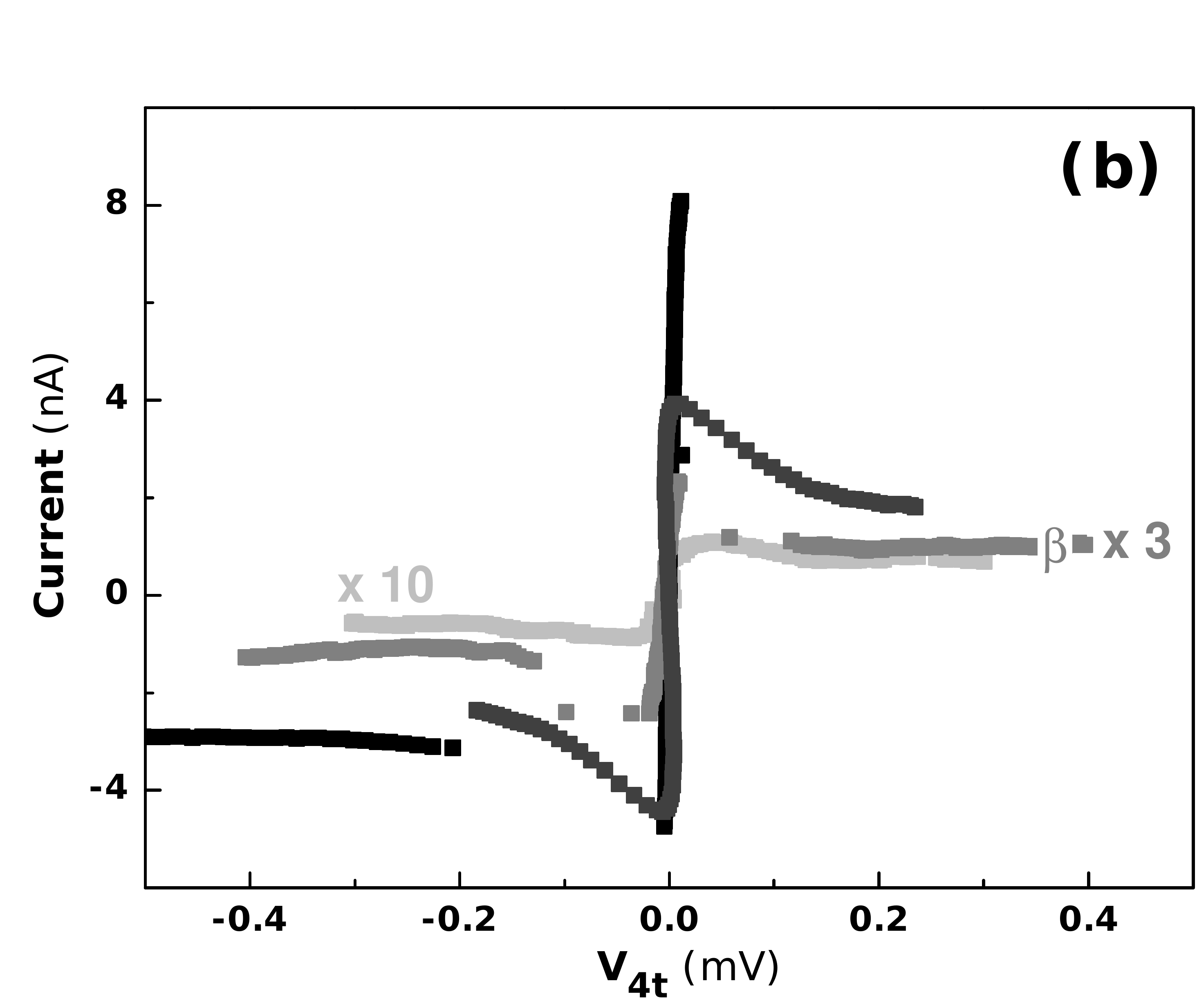}
 \caption{\textbf{(a) Critical currents as a function of the circumference $U$ (right-hand side) and the $\nu_{tot}=1$ sample area $A$ (left-hand side) for $d/l_B\approx$1.6 for samples from wafer $\alpha$ and $\beta$. Fitting yields a linear dependence on the area. See text for further discussions. (b) Corresponding 4-terminal $I/V$ curves. The sample dimensions are (from largest to smallest): Corbino ring: $d_{outer}$=860~$\mu$m, $d_{inner}$=270~$\mu$m; Hall bar type 1: 0.88~$\times$~0.08~mm$^2$; Hall bar type 2: 0.15~$\times$~0.05~mm$^2$.}} \label{fig2}
\end{figure}

Figure \ref{fig2} (a) plots the value of the critical current as a function of the sample circumference $U$ and as a function of the sample's $\nu_{tot}=1$ area $A$ when the effective layer separation $d/l_B$ for all samples is $\approx$~1.6. The data points are labeled with the wafer index $\alpha$ or $\beta$. Our data clearly indicates that the critical current $I_C$ grows as the sample increases in size. When we apply a linear fit in the $log-log$ diagram, we find that the trend is best described by 1.04$log(A)$ for the area and 2.11$log(U)$ for the circumference. Based on these data we propose that the parameter that determines coherent tunneling at $\nu_{tot}=1$ is the sample area with a linear dependence and not the circumference.

Figure \ref{fig2} (b) shows the corresponding 4-terminal $I/V$ curves. Please note that the sample with the largest area (and circumference) displayed a strong hysteresis between up- and downsweeping the applied voltage $V_{2t}$. Subfigure (b) shows the upsweep, where the negative and positive critical currents differ, i.e., $I^-_C<I^+_C$. In a subsequent downsweep, the situation and the absolute values are reversed, i.e., $I^+_C<I^-_C$ and $I^+_C$(upsweep)=$I^-_C$(downsweep). For that reason, the fits in Figure \ref{fig2} (a) ignore the error bar of these data points. 

For comparison we extracted a critical current of 17~pA at $d/l_B\approx$~1.6 from \cite{Spielman01, Spielman04}. This value originates from the dc part of a TSM on a 250~$\times$~250~$\mu$m$^2$ sample. We believe that the disagreement between our data and this value is related to a different single particle tunneling splitting $\Delta_{S,AS}$, which is determined by the height and width of the tunneling barrier, and to the large enhancement of the tunneling amplitude at $\nu_{tot}=1$ \cite{Rossi05}. Their double quantum well is differently designed, with 18 nm GaAs quantum wells separated by a 9.9~nm barrier layer. The center-to-center separation $d$, however, is nearly identical to ours.

Our studies indicate that the initial notion of a length dependence on tunneling as mentioned at the beginning of this chapter is a misinterpretation due to the astonishing properties of the coherent phase. Coherent inter-edge tunneling in a Corbino topology is suppressed because injected electrons do not become part of the $\nu_{tot}=1$ state. In the excitonic condensate picture this can be understood as it is impossible for the condensate to create an interlayer exciton by placing a hole at the site of the injected electron, when the drain lead where the hole has to originate from is far away on the other side of the bulk. True coherent tunneling on the other hand occurs when new interlayer excitons are created or existing ones are annihilated. These processes appear to be bulk phenomena after all. However, as the $\nu_{tot}=1$ phase could break up into domains near the phase boundary \cite{Stern02}, it is possible that certain physical conditions may arise which yield a different dependence on the area (and circumference) than the one presented here. 

Generally, the observation of a critical behavior requires - vaguely speaking - ``sufficiently large samples''. The tangible sample size depends on the underlying design of the double quantum well and the sample structure, but may also be influenced by other unknown factors. More specifically, our studies identify the area rather than the circumference as the determining parameter for coherent tunneling at $\nu_{tot}=1$.


\section{VI. DISCUSSION}

A tunneling experiment in the general sense of charge transfer between two electron reservoirs through a sufficiently thin barrier is an inappropriate interpretation for the peculiar case of the total filling factor one state. Instead, as the two layers are considered to be indistinguishable, correlated interlayer tunneling is a direct signature of interlayer phase coherence. The model of indistinguishable layers also implies that the critical behavior and its dependence on the parameters we discussed in this paper have several implications for magneto-transport experiments performed in the regime of the total filling factor one state \cite{Ezawa08}. It can be shown that the leakage or tunneling current in the counterflow configuration \cite{Kellogg04, Tutuc04} depends on whether the driving current in the system is larger or smaller than the critical current for a given condition \cite{MPI09}, i.e., $d/l_B$, temperature and sample size. The observed gap in magneto-transport on the other hand is only slightly altered. It nevertheless implies that temperature-activation measurements at a total filling factor of one \cite{Wiersma04} require a much more careful interpretation of the gap energy.

What makes matters experimentally complicated is the strong dependence of the critical current $I_C$ on the size of the sample, meaning that for samples of certain dimensions the transport current may already be much larger than the critical current, even when the system is at base temperature. This fact could be able to account for several unsettled observations such as finite dissipation in counterflow experiments for example \cite{Kellogg04}. Future magneto-transport experiments in the regime of the coherent total filling factor one state need to take the relevance of the driving current into account.


In summary, we discussed the relevant parameters necessary for the observation of a critical behavior in the coherent $\nu_{tot}=1$ state. We find a linear dependence of the value of the critical tunneling current $I_C$ on the $\nu_{tot}=1$ area of the sample. $I_C$ will decrease when the system is brought toward the phase boundary by increasing $d/l_B$ or the temperature. It is also rapidly destroyed by moving away from a total filling factor of one.

\section{ACKNOWLEDGMENTS}

We would like to thank both A. H. MacDonald and M. Gilbert for discussions and comments. Our bilayer wafers were grown in collaboration with M. Hauser (Max-Planck Institute, Stuttgart) and H.-P. Tranitz (University of Regensburg). Especially, we would like to acknowledge J. G. S. Lok for his initial and intensive work on our electron double layer systems. This project was supported by the BMBF (German Ministry of Education and Research) under Grant No. 01BM456.\newline


\end{document}